\documentstyle [12pt]{article}
\newcommand{\mincir}{\raise -2.truept\hbox{\rlap{\hbox{$\sim$}}\raise5.truept
\hbox{$<$}\ }}
\newcommand{\magcir}{\raise -2.truept\hbox{\rlap{\hbox{$\sim$}}\raise5.truept
\hbox{$>$}\ }}
\newcommand{\minmag}{\raise-2.truept\hbox{\rlap{\hbox{$<$}}\raise 6.truept\hbox
{$>$}\ }}
\newcommand{\be}{\begin{equation}}
\newcommand{\ee}{\end{equation}}
\newcommand{\ba}{\begin{eqnarray}}
\newcommand{\ea}{\end{eqnarray}}
\newcommand{\brr}{\begin{array}}
\newcommand{\err}{\end{array}}
\newcommand{\bc}{\begin{center}}
\newcommand{\ec}{\end{center}}

\newcommand{\hm}{\,h^{-1}{\mbox{Mpc}}}

\title{{\bf Large--Scale Structure
in Mixed Dark Matter Models with a Non--thermal Volatile Component}}
\author{ {\bf Elena PIERPAOLI}$^{1,2,3}$, {\bf Peter COLES}$^{2}$, \\
 {\bf Silvio BONOMETTO}$^{3,4}$ \& {\bf Stefano BORGANI}$^{1,4}$\\ \\
{\it $^{1}$ SISSA -- International School for Advanced Studies,}\\
{\it via Beirut 2--4, I-34013 Trieste, Italy}\\ \\
{\it $^{2}$ Astronomy Unit, School of Mathematical Sciences,}\\
{\it Queen Mary \& Westfield College, Mile End Road,}\\
{\it London E1 4NS, United Kingdom}\\ \\
{\it $^{3}$INFN -- Sezione di Milano,} \\
{\it Dipartimento di Fisica dell' Universit\`{a} di Milano,}\\
{\it via Celoria 16, I-20133 Milano, Italy}\\ \\
{\it $^{4}$ INFN, Sezione di Perugia,}\\
{\it c/o Dipartimento di Fisica dell' Universit\`{a},}\\
{\it via A. Pascoli, I--06100 Perugia, Italy}}
\date{}

\begin{document}

\maketitle

\vspace{0.5truecm}

\thispagestyle{empty}

\clearpage

\setcounter{page}{1}

\section*{\center Abstract}
We investigate the properties of large--scale structure predicted in a class of
mixed dark matter models in which the volatile  component  (made of particles
with high rms velocity)  derives from the decay of a heavier particle.
Such models based on cold+volatile dark matter (CVDM) differ from the standard
mixture of CDM and massive neutrinos, usually known as CHDM, in that they
involve a  component which has a non--thermal phase space distribution function.
As a consequence, and differently from CHDM models, the value of the redshift at
which volatile particles become non relativistic, $z_{nr}$, can be varied almost
independently of the volatile fraction, $\Omega_X$. We compute transfer
functions for a selection of such models, having $0.1\le \Omega_X\le 0.5$ and
different values of $z_{nr}$. Using linear theory and assuming a scale--free
primordial spectrum, we compare such models with observational constraints on
large--scale galaxy clustering and bulk flows, as well as on the
abundance of galaxy
clusters and high--redshift damped Ly$\alpha$ systems. We find that
these constraints enable us
to discriminate between different $\Omega_X$ and $z_{nr}$;
within the range of the models inspected, those which can be most easily
accommodated by the data
correspond to the parameter choice $\Omega_X\simeq 0.2$ and $z_{nr}
\simeq 2\times 10^4\Omega_X$.
\vspace{2cm}

{\em Subject headings:} Cosmology: large-scale structure of the universe -
galaxies: clusters: general

\clearpage

\section{Introduction}
The outstanding problem of modern cosmology is to understand the formation of
galaxies, clusters of galaxies and large-scale structure in the expanding
universe. A scenario in which these structures form by the growth by
gravitational instability of small--amplitude primordial density fluctuations in
a universe dominated by weakly interacting non--baryonic dark matter has been
the standard framework within this problem has been discussed for many years.
For most purposes the crucial theoretical quantity in the  gravitational
instability picture is the primordial fluctuation spectrum, $P(k)$. In most
theories, density perturbations originate as a quantum phenomenon in the very
early universe with a power--law spectrum of the form $P(k)\propto k^{n}$; $n$
very close to unity  is favoured by many versions of the inflationary universe
picture, as well as by more general considerations. With the discovery of
temperature anisotropies in the Cosmic Microwave Background Radiation (CMBR)  by
the COBE satellite (Smoot et al. 1992), it is possible to fix the  amplitude of
this spectrum in a relatively unambiguous way on very large  scales, of order
1000 Mpc. On smaller scales, the shape of the primordial spectrum is expected to
evolve from its initial power--law form because of the action of various damping
and dissipative processes. During several intermediate stages different
components can have different spectra and, in some models, residual differences
can still be present at the onset of non--linear stages.

In many respects the problem of explaining structure formation in the
gravitational instability picture can  be reduced to that of finding a power
spectrum  whose primordial form  matches the COBE--inferred amplitude on large
scales, and whose evolved form simultaneously matches the  statistical
properties of galaxies and clusters on smaller scales. This has proved to be a
non--trivial task for the Cold Dark Matter (CDM) model, which, at least in its
standard formulation ($\Omega_0=1$, $n=1$, $h=0.5$ and Gaussian adiabatic
fluctuations), is now generally accepted to be ruled out by the data (Wright et
al. 1992; Taylor  \& Rowan--Robinson 1992; Liddle \& Lyth 1993; White,
Efstathiou \& Frenk 1993). The essential problem of this model is that, once it
is normalized to match the measure CMB  temperature anisotropy on large scales,
it has too large a fluctuation amplitude on small scales $\mincir 10\hm$.
Despite its failure, CDM is nevertheless considered as a reference model,
several modifications of it having been suggested in order to remedy
its shortcomings.
Among the ``second generation'' of CDM--based models is the Cold+Hot Dark Matter
(CHDM) model, which assumes that part of the dark matter content is
in the form of
massive neutrinos of mass $m_\nu \sim 10\,$eV (Valdarnini \& Bonometto 1985;
Bonometto \& Valdarnini 1985; Achilli,  Occhionero \& Scaramella 1985; Holtzman
1989; Schaefer, Shafi \& Stecker 1992; Schaefer \& Shafi 1992; Davis, Summers \&
Schlegel 1992; Holtzman \& Primack 1993;  Liddle \& Lyth 1993; Klypin et al.
1993). In this scenario, the small--scale power is suppressed by neutrino
free--streaming by an amount which depends on the hot percentage;
the number of neutrino species participating in
the hot component also plays a role (Primack et al.
1995; Babu, Shafer \& Shafi 1995). This model, with $\sim 20$--30\% to the
density from the hot component appears to provide  a good description of
structure  on a rather large range of scales, although stringent
constraints on the exact amount of the hot
component are provided by the abundance of high--redshift objects (e.g. Ma \&
Bertschinger 1994; Klypin et al. 1995).

In this paper we discuss a scenario  similar to the CHDM picture,  where dark
matter has a cold component (CDM) and a further  volatile component (VDM), which
has a phase space distribution resulting from the decay of a heavier particle
species. A previous paper (Pierpaoli \& Bonometto 1995, hereafter PB95)
calculated the effects of such a particle species upon the evolution of
fluctuations through the period of recombination and up to the present epoch.
Results were presented in PB95
in the form of transfer functions for several examples.
In the following analysis we consider different models with  respect
to those discussed in PB95, so as to provide a
wider sampling of the
cold+volatile dark matter (CVDM) model parameter space.
Furthermore, we go beyond the calculation of the transfer functions and submit
the models to a number of explicit tests, by comparing them with observational
data. The tests performed take into account: (i) the large--scale galaxy
clustering as deduced from the analysis of volume--limited galaxy samples
obtained from redshift catalogues; (ii) the behaviour of bulk velocities
calculated using POTENT; (iii) the observed abundance of galaxy clusters;
(iv) the observed abundance of high--redshift structures traced by damped Ly--$\alpha$
systems. Such tests therefore refer to scales ranging from a hundred of Mpc's
down to a fraction of Mpc.

The plan of the paper is as follows. In Section 2 we introduce the CVDM
models and the corresponding linear power spectra. By only resorting to
linear--theory approaches, we compare in Section 3 such models to
observational data.
In Section
4 we discuss the main results and draw  general conclusions from
our analysis.

\section{The models}
The model we have in mind is one with two dark matter components, as well as
baryonic material. To restrict the possible parameter space we consider only
models with $\Omega_\circ=1$, $H_0=50$ km s$^{-1}$ Mpc$^{-1}$.
As for the
baryonic contribution, we consider the two values $\Omega_b=0.05$ and
$\Omega_b=0.08$, which correspond to the central prediction of standard
nucleosynthesis and to the $95\%$ upper limit allowed by this constraint
(e.g. Reeves 1994) given our chosen value of $H_0$.

In conventional Cold+Hot DM models, the hot component is assumed to be made of
neutrinos with a mass  of a few eV produced in thermal equilibrium in the early
universe. In such a picture the redshift at which the neutrino becomes
non--relativistic is $z_{nr} \simeq 1.4 \times 10^4 (m_\nu / 10\, {\rm eV})$ and
the contribution of $\nu$'s to $\Omega_o$ is $\Omega_\nu \simeq 0.21\, g_\nu
(m_\nu / 10\, {\rm eV})$ ($g_\nu$ is the
 number of neutrino spin states with mass
$m_\nu$, originally in thermal equilibrium). Accordingly $z_{nr} \simeq 6.7
\times 10^4 \Omega_\nu / g_\nu$ and therefore $z_{nr}$ and $\Omega_\nu$ are not
independent, as $g_\nu$ takes only suitable discrete values.

In the scenario we discuss here the VDM particles are produced by the  decay of
a more massive particle and
are consequently
never in thermal equilibrium. For the purposes of this paper, this means
effectively that the parameters $z_{nr}$ and $\Omega_X$, where $X$ denotes the
VDM species, can be varied independently.

It is also important to stress that there exist microphysical models where
decay processes do indeed give rise to a cosmological scenario
of the kind we discuss here. One such model is described
by Bonometto, Gabbiani \& Masiero (1994) and discussed also in PB95.
In this model
both supersymmetry (SUSY) and Peccei--Quinn symmetries hold, and it is
possible that the lightest eigenstate of the ordinary neutral fermion partners
in SUSY theories, called the neutralino, can decay into the supersymmetric
partner of the axion (called the axino).
The non--thermal axinos produced by the decays would be VDM. Plausible values
of the model parameters are $\Omega_X$ in the range 0.1 to 1.0 and $z_{nr}$ in
the range $10^{3}$ to $10^{5}$. There does remain a constraint between
$\Omega_X$ and the number of massless neutrinos which is imposed by primordial
nucleosynthesis considerations (see PB95).
We have mentioned this specific model merely as an illustration;
others can be constructed, but their discussion is beyond the scope
of this paper.

Whatever the origin of a non--thermal volatile component, it has a significant
effect on the nature of the evolved fluctuation spectrum. Our criterion to
select models in the $\Omega_X$--$z_{nr}$ parameter space can be sketched as
follows. Let $g^*$ be the number of helicity states which are allowed by
nucleosynthesis and $g_\nu$ that associated to neutrinos which are present at
the nucleosynthesis epoch. Then $g_X=g^*-g_\nu$ is the number of extra spin
states to be associated to VDM particles. If $z_{nr}$ is the redshift at which
VDM becomes non relativistic, then the limit for $\Omega_X$ reads
\be
\Omega_X \,\mincir \,{z_{nr}\over 10^5}\,g_X
\label{eq:omx}
\ee
(see PB95). For each $\Omega_X$, we choose three different $z_{nr}$ values,
namely $z_{nr}=2\times 10^4 \Omega_X,\,5\times 10^4 \Omega_X$, and
$2\times 10^5 \Omega_X$.

According to Walker et al. (1991),  standard stellar  light
element abundances, combined with nucleosynthesis calculations,
allow up to $g^*=7$, a
value which is also quite close to the $2\sigma$ upper bound provided by Olive
and Scully (1995). Note, however, that Hata et al. (1995) recently gave the
more stringent constraint $g^*<5$ at 95\% C.L., which even conflicts with the
{\em usual} 3 neutrino species. In this paper, we adopt the limit $g^*=7$.

The first value of $z_{nr}$ given above corresponds to $g_X=5$, if
eq.(\ref{eq:omx}) holds as an equality. Accordingly, the number of allowed
neutrino species is  $N_\nu=g_\nu/2=1$. This implies that only one massless
neutrino is present at nucleosynthesis; the remaining two neutrinos are quite
heavy and have been already decayed. Taking $z_{nr}=5\times 10^4\Omega_X$
instead allows $g_X=3$ with $N_\nu=2$, while the choice $z_{nr}=2\times
10^5\Omega_X$ allows one helicity state for VDM particles with $N_\nu=3$.

Models with the first value of $z_{nr}$ would thus be inconsistent with the
Hata et al. (1995) limits. But note also that, contrary to what one might
naively expect,  the results we quote below
for higher values of $z_{nr}$ cannot be used straightforwardly
to accomodate this more stringent constraint because the shape of the
transfer function depends explicitly on the number of massless neutrino
species.

As for $\Omega_X$, we choose values in the interval $0.1 \le \Omega_X\le 0.5$,
with step 0.1. For each value of $\Omega_X$ we allow for  both the above
baryonic fractions. In Table 1 we list the parameters for the resulting 30
models, on which we base the discussion of this paper. Note that, with these
choices for $z_{nr}$ and $\Omega_X$, differ substantially from those
presented in PB95 and probe a more carefully chosen part of the parameter space.

For each model, we follow the evolution of the fluctuations in the baryonic
($\delta_b$), cold ($\delta_c$) and volatile ($\delta_X$) components
through equivalence and recombination epochs
(the algorithm used here has drastically reduced the CPU times with respect to
the one used in PB95; see however
PB95 for its details). The transfer function
is, a priori, different for the various components.
Since we are not interested here in following the evolution of each single
component, in what follows we
define  the overall transfer function as the ratio
\be
T(k)\, =\, {(\delta_b+\delta_c+\delta_X)_{z=0} \over
  (\delta_b+\delta_c+\delta_X)_{z \gg z_{\rm hor}}}\,.
\label{eq:tkdef}
\ee
Accordingly, $T(k)$ is normalized to unity at scales so large that they
enter the horizon after recombination. In the above definition, $z_{\rm hor}$ is
a redshift at which the smallest scale considered is still well above the
horizon scale. Numerical values of $T(k)$ are provided by the parametric expression
\be
T(k)\, =\, \left( 1+\sum_{j=1}^{4} c_j k^{j/2} \right)^{-1},
\label{eq:tk}
\ee
where the $c_j$  generally depend on the redshift, due to the residual
free--streaming of volatile particles. The values of the fitting
parameters at $z=0$ are given in Table 1, with $k$ measured in
Mpc$^{-1}$.  The limiting scale down to which  transfer functions are
computed is 250 kpc.

We will not go into a detailed analysis of the relative behaviour of the
models here, but it is worth pointing out a couple of important trends.
Firstly, at fixed $z_{nr}$ and $N_\nu$, the amount of small scale power
relative to large decreases as $\Omega_X$ is increased, due to the
progressively larger effect of free--streaming. On the other hand, at fixed
$N_\nu$ and $\Omega_X$, the power on small scales increases relative to
large scales as $z_{nr}$ is increased, since a larger $z_{nr}$ corresponds
to smaller velocities at the present time, and a consequent
reduction of the effects of free--streaming.

Assuming that the COBE signal contains a negligible contribution from
gravitational waves (see, e.g., Lidsey \& Coles 1992), the spectrum
normalization can be accomplished by matching the observed COBE fluctuation
spectrum with its predicted form:
\be
C_l=\langle |a_{lm}|^{2}\rangle =
\frac{1}{2\pi} \left({H_0 \over c}\right)^{4} \int_0^{\infty}
P(k) j_{l}^{2}(kx_H)k^{-2} dk,
\ee
where $j_l$ is a spherical Bessel function and $x_H=2c/H_0$.
We take $P(k)=Ak$ (scale--free primordial fluctuations: Harrison
1970; Zel'dovich 1972). The spectrum amplitude $A$ is estimated by
matching the  quadrupole ($l=2$) value of $Q_{\rm rms-PS}=20 \mu$K
(Gorski et al. 1994; Wright et al. 1994), with an approximate 10\%
uncertainty on this value (Gorski, Stompor \& Banday 1995).  The value of
$A$ required to normalize each model in this way is also displayed in
Table 1.

\section{Observational tests}
We concentrate here on comparing analytical predictions of the CVDM models
with observational constraints on large--scale properties of the density
and velocity fields, as well as on abundances of cosmic structures. More
refined investigations involving the use of N--body simulations to
account properly for effects of non--linear clustering, are beyond the
scope of this paper. All the results in the following depend only on the linear
power--spectrum. Results for the CVDM models are also compared to the
predictions of the standard CDM and the CHDM model with $\Omega_\nu =0.3$ for
the fractional density contributed by one massive neutrino species.

\subsection{Power--Spectra}
In order to compare  the power spectra of the CVDM models to that  of the galaxy
distribution, we plot in Figure 1 $P(k)$ for models with $\Omega_X=0.1,0.2,0.3$
and 0.5, as well as for two volume--limited subsamples of the CfA2 and SSRS2
redshift surveys (da Costa et al. 1994). The spectra plotted
are all normalized
to COBE, so that galaxies are assumed to trace exactly the DM distribution
(allowing for linear biasing would just result in a vertical shift of the
curves). Linear--theory redshift space distortions are introduced by multiplying
$P(k)$ by the correction factor $f=1+2/(3b)+1/(5b^2)$ (Kaiser 1987), where we
assume $b=1$ for the biasing factor. Since we plot  linear spectra, we should
take in mind that the comparison with data is  reliable only for $k\mincir
0.2\,(\!\hm)^{-1}$. At smaller scales, non--linear gravitational clustering
makes both the linear spectrum shape  and the treatment for redshift--space
distortions inadequate. In any case,  we expect the net result of
non--linear effects to be an increase of
$P(k)$.   Those models whose linear
$P(k)$ already fall above the data points would
therefore have even harder time if a proper
non--linear treatment were performed; this is the case for
CDM and for all the $\Omega_X=0.1$ CVDM  models. As $\Omega_X$ is increased, we
note that the models with large values of $z_{nr}$ display
more power than the data
for $0.1\mincir k\mincir 0.2\,(\!\hm)^{-1}$. 
This is just
the consequence of the fact that, at larger values of $z_{nr}$ the volatile 
particles have smaller velocities and, therefore, fall into 
the CDM potential wells at an earlier epoch.

In order to provide a more quantitative description of the clustering,
we compute $\sigma_8$, which is defined as the rms fluctuation amplitude,
\be
\sigma_R\,=\,\left[ {1\over 2\pi^2} \int_0^{\infty}
P(k) W^2(kR)k^2 dk\right ] ^{1/2}\,,
\label{eq:sigr}
\ee
within a top--hat window, $W(kR)=3(\sin{kR}-kR\cos{kR})/(kR)^3$, of radius
$R=8\hm$. The $\sigma_8$ values for all the models are listed in Table 2 and are
compared with that, $\sigma_8=0.90\pm 0.05$, reported by Loveday et al. (1995)
for Stromlo--APM galaxies in real space.  It turns out that models with
$\Omega_X= 0.1$ or with large $z_{nr}$ have the unpleasant feature of implying a
substantially antibiased galaxy distribution, in some cases comparable to that
of CDM. A constraint on $\sigma_8$   for the DM distribution comes from the
cluster abundance, which suggests  $\sigma_8\simeq 0.6$ for $\Omega_0=1$ models
(e.g. White, Efstathiou \&  Frenk 1993; see below for a more detailed
discussion of the cluster  abundance predictions for our models). We note that
only small variations  of $\sigma_8$ are obtained by increasing the baryonic
contribution from 5\%  to 8\%.

Smaller values of $\sigma_8$ may be
allowed by the COBE normalisation if one tilts the primordial spectral index
to $n<1$, also possibly allowing for the presence of some tensor
contribution in the CMB temperature anisotropies. For instance, taking
$n=0.9$ for the model 4 ($\Omega_X=0.2$, $z_{nr}=4\times 10^3$) one gets
$\sigma_8=0.70$.

As a further characterization of the power--spectrum shape,
we computed  the parameter $\Gamma$ defined by
$\Gamma=0.5(3.4\sigma_{25}/0.95\sigma_8)^{-1/0.3}$ (Wright et al. 1992;
Efstathiou, Bond \& White 1992);
$\Gamma\simeq 0.25$ is required by the data.
According to the values reported in Table
2, we confirm the visual impression obtained from Figure 1: models with
$\Omega_X=0.1$ have a too large $\Gamma$ (they are too similar to
CDM), while models with a larger amount of volatile component fare better,
unless one takes a large value of $z_{nr}$.

\subsection{Bulk Velocities}
The rms bulk velocity, $V_{\rm bulk}(R)$, is defined as the rms
matter velocity after smoothing over a volume of size $R$.
 For $\Omega_\circ=1$ it is connected to the
power--spectrum according to
\be
V_{\rm bulk}^2(R)\, =\, {H_0^2\over 2\pi^{2}} \int_0^{\infty}
P(k) W^{2}(kR) dk,
\label{eq:vbu}
\ee
where $W(kR)$ is the window function specifying the shape of the
smoothing volume. By
comparing eqs.(\ref{eq:sigr}) and (\ref{eq:vbu}), it is clear that
$V_{\rm bulk}(R)$ gives more weight to long wavelength modes than
than $\sigma_R$.
Therefore, we expect bulk velocities on large scales to depend only on the
$Q_{rms-PS}$ normalization and not on the profile of the transfer function.

Reliable $V_{\rm bulk}$ data for top--hat spheres centered on the Local Group on
scales of few tens of Mpcs are provided by the POTENT reconstruction method
(e.g. Bertschinger et al. 1990; see also Dekel 1994, and references therein).
In Figure 2 we compare our model predictions with the latest POTENT data
(courtesy of A. Dekel). In order to account for the velocity smoothing
procedure
in the reconstruction method, we convolved the power--spectrum in
eq.(\ref{eq:vbu}) with a Gaussian filter of radius $R_f=12\hm$.
As expected, any difference between models at large scales
is negligible and all of them are in remarkable agreement with data. On
smaller scales ($\mincir 40\hm$) the predicted $V_{\rm bulk}$ values tend to be
larger than the observational one. In this respect, models with
$\Omega_X\ge 0.4$ perform better, thanks to the steep $P(k)$ profile at
large $k$, although this is unfortunately
inconsistent with galaxy
clustering data. On the other hand, independent estimates of bulk flows
(e.g. da Costa et al. 1995) agree with the POTENT one only on scales $R
\simeq 50$--60$\hm$. Therefore, we do not regard this marginal discrepancy
as a serious problem for any of the models we have considered.

\subsection{Cluster abundance}
According to the standard Press \& Schechter (1974) approach, the number density
of collapsed structures arising from Gaussian initial fluctuations and having
mass larger than $M$ is given by
\be
N(>M)\,=\,\int_M^\infty n(M')\,dM'\,.
\label{eq:abo}
\ee
Here $n(M)\,dM$ is the number density of objects with mass in the range
$[M,M+dM]$ and is related to the power--spectrum according to
\be
n(M)\,dM~=~{1\over \sqrt{2\pi}}\,{\delta_c\over f}
\int_R^\infty{\eta_R\over \sigma_R}\,\exp\left(-{\delta_c^2\over 2\sigma^2_R
}\right)\,{dR\over R^2}\,,
\label{eq:ps}
\ee
where
\be
\eta_R \, = \, {1\over 2\pi^2 \sigma^2_R}\int k^4\,P(k)\,
{dW^2(kR)\over d(kR)}\,{dk\over kR}\,.
\label{eq:etsi}
\ee
In the above expressions, we assume that the mass scale $M$ is related to  the
length scale $R$ according to
$M=f\bar\rho R^3$, with $f$ the ``form factor", which is
specified by the shape of the filter $W$ and $\bar \rho$ the average
density. For the Gaussian window, that we assume here, it is $f=(2\pi)^{3/2}$,
while $f=4\pi/3$ for a top--hat window. The parameter $\delta_c$ is the critical
density contrast, which represents the threshold value for a fluctuation to turn
into an observable object, if  evolved to the present time by linear theory. For
a top--hat spherical collapse one has $\delta_c=1.68$, but the inclusion of
non--linear effects, as well as aspherical collapse, may lead to a lower value
of  $\delta_c$. For example, Klypin \& Rhee (1994; KR94 hereafter) found that
the  cluster mass function in their CHDM N--body simulations is well fit by
eq.(\ref{eq:ps}) by taking $\delta_c=1.5$ for the Gaussian window. In order
to account for the rather poor knowledge of $\delta_c$, we prefer to compute
$N(>M)$ for different values of this parameter in the range $[1.4,1.7]$.

The results of this analysis are reported in Figure 3, where  we also compare
them with observational results. Values  of $N(>M)$ for $\delta_c=1.5$ are also
listed in Table 2. Following White et al.  (1993), we take $M=4.2\times
10^{14}h^{-1}M_\odot$ for the limiting mass at which  to estimate the mass
function. We prefer not to  consider a larger value, $\sim
10^{15}h^{-1}M_\odot$, since this would  correspond to the exponential tail of
the cluster mass function (e.g.  Bahcall \& Cen 1992) and, as a consequence,
large variations in the cluster abundance  would be associated with
uncertainties in the
cluster mass estimates. The dashed band in Figure 3 corresponds to the range
between the  observational result of White et al. (1993; lower limit) based on
$X$--ray  data, and that of Biviano et al. (1993; upper limit) based on
velocity dispersions. We note that realistic observational uncertainties  are
probably larger than the difference between such two results. They may  be due
to systematic effects, related to assumptions used
to connect $X$--ray  temperature
and  DM potential profiles, or to biases in estimating cluster masses from
internal velocities under the virial assumption.

Even bearing such warnings in mind, it seems difficult to reconcile with the
data those models which overproduce clusters by one order of magnitude or
more, for any value of $\delta_c$ value. This is the case for
$\Omega_X=0.1$ and, in general, for those models having a large value of
$z_{nr}$. Even though taking the larger baryonic fraction decreases the
cluster abundance, its effect is nevertheless only marginal for those
models which have an exceedingly large $N(>M)$.

It is not clear whether such large discrepancies may be overcome on the  ground
of observational biases. For instance, let us consider the model 5
($\Omega_X=0.2, z_{nr}=10^4$) as a case providing a large $N(>M)$. If we allow
for an  underestimate of cluster masses by a factor 2 (i.e., $M=8.4\times
10^{14}h^{-1}M_\odot$; see, however, Evrard, Metzler \& Navarro 1995, for
arguments in favour of precise mass determinations from $X$--ray data)  and take
$\delta_c=1.5$, it would give  $N(>M)\simeq 1.2\times 10^{-5}(\!\hm)^{-3}$,
which is still  quite far from the observational band.

On the other hand, we do not believe that the observational situation is  clear
enough to rule out at a large confidence level  models which are discrepant by a
factor 2--3  with respect to the reported abundances. Even adopting  such a
rather conservative  position, it is fair to say that only models with
$\Omega_X\ge 0.2$ and  low $z_{nr}$ are not excluded by this constraint. In this
respect, the availability of more and more determinations of
cluster masses based
on the independent technique of weak gravitational lensing (e.g. Squires et al.
1995) will be extremely welcome.

\subsection{High--redshift behaviour}
A further constraint on power spectra comes from observations of
high--redshift objects. The most reliable such constraint concerns the
abundance of damped Ly--$\alpha$ systems (DLAS). These are observed as
wide absorbtion troughs in quasar spectra, due to a high HI column density
($\ge 10^{20}$ cm$^{-2}$). The fact that at $z\sim 3$ the fractional
density of HI gas associated with DLAS is comparable to that contributed by
visible matter in nearby galaxies, suggests that DLAS actually trace a
population of collapsed protogalactic objects (see Wolfe 1993, for a
comprehensive review). Lanzetta et al. (1995) and Wolfe et al. (1995)
presented data on DLAS up to the redshift $z\simeq 3.5$, while the recent
compilation by Storrie--Lombardi et al. (1995) pushed this limit to
$z\simeq 4.25$. Based on these data, the latter authors claimed the
first detection of a turnover in the fractional density, $\Omega_g$,
of neutral gas belonging to the absorbing systems at high redshift.

Several authors recognized DLAS as a powerful test for DM models, based on a
linear  theory approach, Subramanian \& Padmanabhan (1994),
Kauffman \& Charlot (1994), Mo \& Miralda--Escud\'e
(1994) and Ma \& Bertschinger (1994) concluded that the standard CHDM scenario
with $\Omega_\nu=0.3$ is not able to generate enough collapsed structures
at $z\magcir 3$, due to the lack of power on galactic scales. However, either
lowering $\Omega_\nu$ to about 0.2 (Klypin et al. 1995) or `blueing' the
primordial spectrum, $P_i(k)\propto k^n$ to $n\simeq 1.2$ (Borgani et al.
1995) keeps CHDM into a better agreement with data. Katz et al. (1995)
resorted to numerical simulations of DLAS and found that even the
CDM model with a normalization as low as $\sigma_8=0.7$ satisfies the DLAS
constraint.

In  order to connect model predictions from linear theory with observations,
let
\be
\Omega_{\rm coll}(M,z)\,=\,{\mbox{\rm erfc}}
\left({\delta_c \over \sqrt 2 \sigma(M,z)}
\right)\,,
\label{eq:omcol}
\ee
be defined as the fractional density contributed at the redshift $z$ by
collapsed structures of mass larger than $M$. Accordingly, it is
$\Omega_g=\alpha_g \Omega_b \Omega_{\rm coll}$, where $\alpha_g$ is the
fraction of HI gas which is involved in the absorbers.

One expects the value of $\alpha_g$, to decrease well below unity
at low redshift, due to gas consumption into stars.  Note that varying this
number turns into a linear rescaling of $\Omega_g$. Since we assume here that
$\alpha_g=1$, we compare data and CVDM predictions at the highest redshift
allowed by the data,  $z=4.25$  (Storrie--Lombardi et al. 1995). We estimate
$\sigma(M,z=4.25)$ for a Gaussian window, by explicitly computing the transfer
function at that redshift, so as to take into account effects of residual
free--streaming of the volatile component.

As for the value of the critical density contrast $\delta_c$, results based
on N--body approaches by Efstathiou \& Rees (1988) and by Klypin et al. (1995)
suggests that $1.3\mincir \delta_c\mincir 1.5$ for a Gaussian window, 
while Ma \& Bertschinger
(1994) found indications for $\delta_c\simeq 1.7$--1.8 for a top--hat
window. In the following we will
report results for $\delta_c=1.5$ and gaussian window; while the effect of 
varying  this parameter is discussed in more detail by Borgani et al. (1995). 
Lacey \& Cole (1994) have realized a detailed test of the Press--Schechter 
theory against scale--free N--body simulations, by checking the effects
of varying the window and the halo identification method. As a result, they
found that $\delta_c \simeq 1.3$ and $\delta_c\simeq 1.8$ are in general 
adequate to describe the halo mass function for Gaussian and top--hat window, 
respectively. In any case, we
verified that rather similar results are obtained either using $\delta_c=1.5$
with a Gaussian window, or $\delta_c=1.7$ with a top--hat window.

The results of our analysis are reported in Figure 4, where we plot the neutral
gas fraction associated to DLAS, $\Omega_g$, for all the models with
$\Omega_b=0.05$ and compare them with the observational data. In the light of
all the above--mentioned uncertainties in realising such a comparison, we prefer
here to adopt a conservative approach and to consider in this comparison the
result of Storrie--Lombardi et al. (1995; $\Omega_g =2.2\pm 0.5$ at $z=4.25$ for
$\Omega_0=1$ and $h=0.5$) as a lower bound. Consistently, the dashed areas in
Figure 4 are delimitated by the above $1\sigma$ lower limit. Only model falling
below this limit are ruled out. The effect of varying the limiting mass of the
protogalaxy hosting DLAS by an order of magnitude may be judged by comparing
open and filled dots, which correspond to $M=5\times 10^9 h^{-1}M_\odot$ and
$M=5\times 10^{10}  h^{-1}M_\odot$, respectively. Numerical values of $\Omega_g$
for all the models are reported in column 5 of Table 2, where also results for
CDM and CHDM are given.

As expected, taking lower $\Omega_X$ and larger $z_{nr}$ make easier the
agreement with data. All the models with $\Omega_X\le 0.2$ are able to pass the
DLAS test, while larger volatile fractional densities are only allowed for large
$z_{nr}$, which however turns into a wrong power--spectrum shape.

Although several models can clearly be ruled out already at this level  on the
ground of DLAS data, nevertheless it is clear that more precise conclusions can
only be reached with a better knowledge of the variables entering in the
Press--Schechter prediction for $\Omega_g$ (i.e. the mass $M$,
and the parameters $\delta_c$ and
$\alpha_g$). A more accurate definition of what is a DLAS in a given DM model
can only be achieved with numerical simulations involving hydrodynamics
(Katz et al. 1995), which would be able to trace the history of galaxy formation.
As for observations, the possibility that systematic biases may affect the final
results have been recently suggested. For instance, Bartelmann \& Loeb (1995)
have recently pointed out that amplification biases due to gravitational lensing
of QSOs by DLAS could led to an overestimate of $\Omega_g$, by an amount which
however decreases with redshift. Fall \& Pei (1995) argued that dust
obscuration may act in the opposite direction so as to bias downwards the
estimated $\Omega_g$. Verifying the actual relevance of such effects surely
requires a substantial investment of observational and theoretical effort.

\section{Discussion}
The results we have presented demonstrate that the CVDM hypothesis
yields potentially interesting models of structure formation.
The aim of this paper is to show that
rather slight changes in the parameters of volatile dark matter can make a
significant difference to the transferred power spectrum.
This contrasts
with the case of a cold component, where the physical properties of the
candidate particle do not really matter at all, in that the physical origin of
the hot particles may leave a detectable imprint in the clustering pattern.
In this context it is important to verify up to which point the shape of the
distribution function  causes
differences compared to the
standard scenario based on relic thermal neutrinos.

The ability to change $z_{nr}$ almost independently of $\Omega_X$ is
especially significant in this respect: the power spectra we have obtained
display considerable variations at a fixed value of $\Omega_X$. Indeed, although
the CVDM class of models involves one more parameter than is the case for
CHDM, we
have shown that observational data nevertheless
allow us to put rather stringent constraints on
the permitted values of $z_{nr}$ and $\Omega_X$, even at the level of
linear--theory. The most stringent of these constraints comes from the
simultaneous requirement for a model to satisfy the observed abundance of
high--redshift DLAS and of galaxy clusters. As for DLAS, the rather large value
of the HI gas fraction involved in the absorbing systems, $\Omega_g$, implies a
substantial amount of power on galaxy scales, so as to favour models with
$\Omega_X\mincir 0.2$. A larger volatile component would be allowed only
resorting to a high value of $z_{nr}\simeq 2\times 10^5 \Omega_X$ (cf. Figure
4). On the other hand, models with small $\Omega_X$ and/or large $z_{nr}$
behave too much like the standard CDM model, drastically
overproducing clusters (cf. Figure 3).

Therefore, the overall result would be that models with $z_{nr} \magcir 5\times
10^4 \Omega_X$ have a hard time, quite independently of $\Omega_X$.
Among the models inspected,
the only model which passes all the tests, or at least which can not be
confidently ruled out, is the one with $\Omega_X=0.2$ and $z_{nr}= 4\times
10^3$. It is worth recalling, however, that such a model with low $z_{nr}$
requires that volatile particles occupy at least 5 helicity states [cf.
eq.(1)]. We recall that this can accommodated only if {\bf (a)} $g^*=7$ is
allowed by standard nucleosynthesis and {\bf (b)} two neutrino species are
sufficiently massive that they have already decayed at the nucleosynthesis
epoch.

An alternative possibility, holding if the physics of the decay is quite
different from the axino model suggested in PB95, is that the decay itself
takes place after the nucleosynthesis epoch. This would make low $z_{nr}$
models compatible with all $N_\nu$. It must, however, be
remembered that
changing $N_\nu$ itself
causes an alteration of the transfer function, and
a straightforward extrapolation of the above results to greater $N_\nu$
values is not allowed. Changing the relation between $N_\nu$ and $z_{nr}$
opens the way to inspecting different models and, in this context,
we should also bear in mind that our analysis has been
based on assuming a scale--free primordial spectrum, while  variations around
this model are allowed by some classes of inflationary schemes. For instance,
taking $P_i(k)\propto k^n$ with $n<1$ (Adams et al. 1993; Liddle \& Lyth
1993 and references therein) decreases the amount of power on the cluster mass
scale, so as to alleviate the problem of cluster overproduction displayed by
``colder" models. However, the amount of this tilt can not be too large, in
order not to conflict with CMB (Bennet et al. 1994) and large--scale
peculiar motions constraints (Tormen et al. 1993). On the other hand, the
case of ``antitilting'', with $n\simeq 1.2$,
has been recently advocated to alleviate
some of the problems of the CHDM scenario (Dvali, Shafi \& Schaefer 1994;
Lucchin et al. 1995). However, the subsequent increase of power on small scales
goes in the undesired direction as far as the cluster abundance is concerned
(Pogosyan \& Starobinsky 1995; Borgani et al. 1995).

As a final remark, we should stress that the analysis presented in this paper is
only preliminary and is entirely based on linear calculations.
In order to be more definitive we would like to extend it  in two main
directions. Firstly to calculate more detailed properties of the CMBR
fluctuations they produce: we anticipate a rather different signature on angular
scales around a degree than in the standard CHDM models.
Furthermore, we
would also like to study the non--linear evolution of
some of
these model
by performing numerical calculations using N--body and other procedures.

\section*{Acknowledgements}
We wish to thank Michael Vogeley for providing us with
the electronic version of
the CfA2+SSRS2 power spectrum. The authors acknowledge the use of the QMW
starlink facilities during this work. Elena Pierpaoli acknowledges
financial support under the EC Human Capital \& Mobility Network `Large
Scale Structure in the Universe: Evolution and Statistics' (Contract No:
ERBCHRX-CT93--0129).  Stefano Borgani has been partially supported by
Italian MURST. Peter Coles is a PPARC Advanced Research Fellow. We are
extremely grateful to  Carlton Baugh for his help with a preliminary
version of this paper.

\newpage
\parindent 0pt
\section*{References}

Achilli S., Occhionero F., Scaramella R., 1985, ApJ, 299, 577

Adams F.C., Bond J.R., Freese K., Freeman J.A., Olinto A.V., 1993, Phys. Rev. D,
47, 426

Babu K.S., Schaefer R.K., Shafi Q., 1995, preprint ASTRO--PH/9507006

Bahcall N.A., Cen R, 1992, ApJ, 398, L1

Bartelmann M., Loeb A., 1995, ApJ, submitted, preprint ASTRO--PH/9505078

Bahcall N.A., Cen R., 1992, ApJ, 398, L81

Bennet C.L., et al. 1994, ApJ, 430, 423

Bertschinger E., Dekel A., Faber S.M., Dressler A.,
Burstein D., 1990, ApJ, 364, 370

Biviano A., Girardi M., Giuricin G., Mardirossian F., Mezzetti M., 1993, ApJ,
41, L13

Bonometto S., Gabbiani F, Masiero A., 1994, Phys. Rev. D., 49,
3918

Bonometto S., Valdarnini R., 1985, ApJ, 299, L71

Borgani S., Lucchin F., Matarrese S., Moscardini L., 1995, MNRAS, in press,
preprint ASTRO--PH/9506003

da Costa L.N., Vogeley M.S., Geller M.J., Huchra J.P., Park C., 1994,
ApJ, 437, L1

da Costa et al., 1995, in
eds. Balkowski C., Maurogordato S., Tao C. \& Tr\^an Thanh V\^an J.,
Proc. of the Moriond Astrophysics Meeting on Clustering in the Universe,
in press

Davis M., Summers F.J., Schlegel D., 1992, Nat, 359, 393

Dekel A., 1994, ARA\&A, 32, 371

Dvali G., Shafi Q., Schaefer R.K., 1994, Phys. Rev. Lett. 73, 1886

Efstathiou G., Bond J.R., White S.D.M., 1992, MNRAS, 258, 1p

Efstathiou G., Rees M.J., 1988, MNRAS, 230, 5p

Evrard A.E., Metzler C.A., Navarro J.F., 1995, ApJ, preprint
ASTRO--PH/9510058

Fall S.M., Pei Y.C., 1995, in QSO absorbtion Lines. Springer Verlag, Berlin,
in press

Gorski K.M. et al., 1994, ApJ, 430, L89

Gorski K.M., Stompor R., Banday A.J., 1995, preprint ASTRO--PH/9502033

Holtzman J.A., 1989, ApJS, 71, 1

Holtzman J.A., Primack J.R., 1993, ApJ, 405, 428

Kaiser N., 1987, MNRAS, 227, 1

Katz N., Weinberg D.H., Hernquist L., Miralda--Escud\'e J., 1995, preprint
ASTRO--PH/95109106

Kauffmann G., Charlot S., 1994, ApJ, 430, L97

Klypin A., Holtzman J., Primack J.R., Reg\"os E., 1993, ApJ, 415, 1

Klypin A., Borgani S., Holtzman J., Primack J.R., 1995, ApJ, 444, 1

Lacey C., Cole S., 1994, MNRAS, 271, 676

Lanzetta K.M, Wolfe A., Turnshek D.A., 1995, ApJ, 440, 435

Liddle A.R., Lyth D.H., 1993, Phys. Rep., 231, 1

Lidsey J.E., Coles P., 1992, MNRAS, 258, 57p

Loveday J., Efstathiou G., Maddox S.J., Peterson B.A., 1995, ApJ, 442,
457

Lucchin F., Colafrancesco S., de Gasperis G., Matarrese S., Mei S., Mollerach
S., Moscardini L., Vittorio N., 1995, ApJ, in press

Ma C.P., Bertschinger E., 1994, ApJ, 434, L5

Mo H.J., Miralda--Escud\'e J., 1994, ApJ, 430, L25

Olive K.A., Scully S.T., 1995, preprint ASTRO--PH/9506131

Peacock J.A., Dodds S.J., 1994, MNRAS, 267, 1020

Pierpaoli E., Bonometto S.A., 1995, A\& A, 300, 13

Primack J.R., Holtzman J., Klypin A., Caldwell D.O., 1995, Phys. Rev. Lett.,
74, 2160

Pogosyan D. Yu., Starobinsky A.A., 1995, ApJ, 447, 465

Press W.H., Schechter P.L., 1974, ApJ, 187, 425

Reeves H., 1994, Rev. Mode. Phys., 66, 193

Schaefer R.K., Shafi Q., 1992, Nat, 359, 199

Schaefer R.K., Shafi Q. Stecker F., 1989, ApJ, 347, 575

Smoot G.F. et al., 1992, ApJ, 396, L1

Squires G., Kaiser N., Babul A., Fahlman G., Woods D., Neumann D.M.,
B\"ohringer H., 1995, preprint ASTRO--PH/9507008

Storrie--Lombardi L.J., McMahon R.G., Irwin M.J., Hazard C., 1995, in
Proceedings of the ESO Workshop on QSO Absorbtion Lines, preprint
ASTRO--PH/9503089

Subramanian K., Padmanabhan T., 1994, preprint ASTRO--PH/9402006

Taylor A.N., Rowan--Robinson M., 1992, Nat, 359, 396

Tormen G., Moscardini L., Lucchin F., Matarrese S., 1993, ApJ, 411, 16

Valdarnini R., Bonometto S., 1985, A\& A, 146, 235

van Dalen A., Schaefer R.K., 1992, ApJ, 398, 33

Walker T.P., Steigman G., Schramm D.N., Olive K.A., Kang H., 1991, ApJ, 376, 51

White S.D.M., Efstathiou G., Frenk C.S., 1993, MNRAS, 262, 1023

Wolfe A., 1993, in Relativistic Astrophysics and Particle Cosmology, eds. C.W.
Ackerlof, M.A. Srednicki, (New York: New York Acad. Sci.), p.281

Wolfe A.M., Lanzetta K.M., Foltz C.B., Chaffee F.H., 1995, preprint

Wright E.L. et al., 1992, ApJ, 396, L9

Wright E.L., Smoot G.F., Bennett C.L., Lubin P.M., 1994, ApJ, 436, 443

\newpage
\section*{Figure Captions}
{\bf Figure 1.} Comparison between observational and CVDM linear power--spectra
in redshift space. Open and filled triangles are the power--spectrum for two
volume limited subsamples of the combined CfA2+SSRS2 survey (da Costa et al.
1994). Each panel refers to a fixed value of the volatile fraction
$\Omega_X$, while the
dotted, short--dashed and long--dashed curves correspond to
different $z_{nr}$. For reference, we also plot the CDM (solid curves).
All the models are for $\Omega_b=0.05$.

\noindent
{\bf Figure 2.} Comparison between the POTENT bulk flow (filled dots)
and that provided by the CVDM models. Different panels are for
different $\Omega_X$ values. In each panel, different curves refer to the
different values for $z_{nr}$.

\noindent
{\bf Figure 3.} The abundance of galaxy clusters with $M>4.2\times
10^{14}h^{-1}M_\odot$. The shaded area is delimited by the observational results
by Biviano et al. (1993; upper limit) and by White et al. (1993; lower limit).
In each panel, corresponding to different $\Omega_X$ values, dotted,
short--dashed and long--dashed curves are for the three different values of
$z_{nr}$. Each pair of curves correspond to the two values of
$\Omega_b$, the lower one being for $\Omega_b=0.08$ and the
higher one for $\Omega_b=0.05$.

\noindent
{\bf Figure 4.} The fractional density of neutral gas involved in DLAS at
redshift $z=4.25$. The shaded area is the observational constraint and
is delimited from below by the 1$\sigma$ lower limit by Storrie--Lombardi et al.
(1995). Each panel refers to a fixed $\Omega_X$ value and reports the predicted
$\Omega_g$ as a function of $z_{nr}$. Filled and open dots correspond to
$M=5\times 10^9\,h^{-1}M_\odot$ and $M=5\times 10^{10}h^{-1}M_\odot$ for the
limiting mass of the protostructures hosting DLAS. We assume a Gaussian window
with $\delta_c=1.5$ and $\alpha_g=1$ for the HI gas fraction involved in the
absorber.

\newpage
\thispagestyle{empty}
\begin{table}[h]
\centering
\caption[]{Model parameters and power spectra. Column 2: number of massless
neutrino species; Column 3: volatile fractional density; Column 3: redshift at
which the  volatile component  becomes non-relativistic (in units of $10^{4}$);
Columns 5 to 9: fitting parameters of the transfer functions [see
eq.(\ref{eq:tk})].}
\tabcolsep 6pt
\begin{tabular}{ccccccccc}  \\ \hline \hline
Model & ${\bf N_\nu}$ & ${\bf \Omega_X}$ & ${\bf z_{nr}/10^{4}}$ & $A\,(10^7)$ &
${\bf c_1}$ & ${\bf c_2}$ & ${\bf c_3}$ & ${\bf c_4}$ \\ \hline \\
\multicolumn{9}{c}{\bf $\Omega_b=0.05$} \\
1 & 1 & 0.1 & 0.2 & 1.300 & --0.6133E+00 &  0.1494E+02 &  0.1220E+03 &  0.6733E+02 \\
2 & 1 & 0.1 & 0.5 & 1.306 & --0.4284E+00 &  0.1212E+02 &  0.9742E+02 &  0.6607E+02 \\
3 & 3 & 0.1 & 2.0 & 1.311 & --0.3665E+00 &  0.1542E+02 &  0.7916E+02 &  0.8917E+02 \\
4 & 1 & 0.2 & 0.4 & 1.318 &  0.5110E--01 & --0.1016E+01 &  0.1924E+03 &  0.1559E+03 \\
5 & 1 & 0.2 & 1.0 & 1.319 &  0.1032E--01 &  0.4255E+01 &  0.1113E+03 &  0.1629E+03 \\
6 & 3 & 0.2 & 4.0 & 1.327 &  0.4886E--02 &  0.1624E+02 &  0.3607E+02 &  0.1997E+03 \\
7 & 1 & 0.3 & 0.6 & 1.354 &  0.1263E+01 & --0.2246E+02 &  0.2301E+03 &  0.3903E+03 \\
8 & 1 & 0.3 & 1.5 & 1.303 & --0.5240E+00 &  0.1509E+02 &  0.2086E+02 &  0.4061E+03 \\
9 & 3 & 0.3 & 6.0 & 1.266 & --0.1997E+01 &  0.5180E+02 & --0.1305E+03 &  0.4194E+03 \\
10& 1 & 0.4 & 0.8 & 1.325 &  0.1052E+00 &  0.8104E+01 & --0.1939E+02 &  0.1053E+04 \\
11& 1 & 0.4 & 2.0 & 1.248 & --0.2475E+01 &  0.5754E+02 & --0.2466E+03 &  0.9205E+03 \\
12& 3 & 0.4 & 8.0 & 1.168 & --0.5488E+01 &  0.1240E+03 & --0.4753E+03 &  0.8259E+03 \\
13& 1 & 0.5 & 1.0 & 1.249 & --0.3115E+01 &  0.1033E+03 & --0.7761E+03 &  0.2781E+04 \\
14& 1 & 0.5 & 2.5 & 1.107 & --0.7916E+01 &  0.1846E+03 & --0.9825E+03 &  0.2091E+04 \\
15& 3 & 0.5 & 10.0& 1.063 & --0.9233E+01 &  0.1943E+03 & --0.8273E+03 &  0.1325E+04 \\ \\
\multicolumn{9}{c}{\bf $\Omega_b=0.08$} \\
16 & 1 & 0.1 & 0.2 &    1.323  &--0.1009E--01  & 0.8960E+01 &  0.1418E+03 &  0.6768E+02\\
17 & 1 & 0.1 & 0.5 &    1.319  &--0.2039E--01  & 0.5736E+01 &  0.1202E+03 &  0.6457E+02\\
18 & 3 & 0.1 & 2.0 &    1.318  &--0.6469E--01  & 0.6622E+01 &  0.1021E+03 &  0.8145E+02\\
19 & 1 & 0.2 & 0.4 &    1.308  &--0.4181E--01  &--0.9723E+01 &  0.2449E+03 &  0.1497E+03\\
20 & 1 & 0.2 & 1.0 &    1.348  & 0.9112E+00  &--0.1115E+02 &  0.1681E+03 &  0.1530E+03\\
21 & 3 & 0.2 & 4.0 &    1.317  &--0.1257E+00  & 0.1017E+02 &  0.5465E+02 &  0.1926E+03\\
22 & 1 & 0.3 & 0.6 &    1.443  & 0.3845E+01  &--0.5984E+02 &  0.3617E+03 &  0.3668E+03\\
23 & 1 & 0.3 & 1.5 &    1.346  & 0.7783E+00  &--0.3884E+01 &  0.8075E+02 &  0.3960E+03\\
24 & 3 & 0.3 & 6.0 &    1.203  &--0.3672E+01  & 0.6458E+02 & --0.1643E+03 &  0.4436E+03\\
25 & 1 & 0.4 & 0.8 &    1.392  & 0.2294E+01  &--0.3311E+02 &  0.1601E+03 &  0.1005E+04\\
26 & 1 & 0.4 & 2.0 &    1.244  &--0.2619E+01  & 0.6079E+02 & --0.2866E+03 &  0.1030E+04\\
27 & 3 & 0.4 & 8.0 &    1.159  &--0.5641E+01  & 0.1198E+03 & --0.4603E+03 &  0.8270E+03\\
28 & 1 & 0.5 & 1.0 &    1.279  &--0.2064E+01  & 0.8336E+02 & --0.6956E+03 &  0.2808E+04\\
29 & 1 & 0.5 & 2.5 &    1.148  &--0.6521E+01  & 0.1636E+03 & --0.1016E+04 &  0.2456E+04\\
30 & 3 & 0.5 & 10.0&    0.925  &--0.1341E+02  & 0.2297E+03 & --0.9345E+03 &  0.1439E+04 \\
\hline
\end{tabular}
\end{table}

\newpage
\thispagestyle{empty}

\begin{table}[h]
\centering
\caption[]{Statistical properties of large--scale structure.
Column 2: r.m.s. fluctuations within a top--hat sphere of $8\hm$ radius. The
observational result refers to the APM galaxy distribution in real space
(Loveday et al. 1995).
Column 3: `extra power' parameter (see text); observational result from Peacock
\& Dodds (1994).
Column 4: number density of clusters with mass larger than $M=4.2\times
10^{14}h^{-1}M_\odot$ (in units of $10^{-6}(\hm)^{-3}$).
Lower and upper values for the observational result are from White,
Efstathiou \& Frenk (1993) and Biviano et al. (1993), respectively.
Column 5: fractional density of neutral gas within collapsed structures of
mass $5\,10^{10}h^{-1}M_\odot$ at redshift $z=4.25$, in units of $10^{-3}$;
the Gaussian window with $\delta_c=1.5$ is assumed; observational result
from Storrie--Lombardi et al. (1995).}
\tabcolsep 6pt
\begin{tabular}{ccccc}  \\ \hline \hline
Model & ${\bf \sigma_8}$ & ${\bf \Gamma}$ &
$N(>M)$ & ${\bf \Omega_g}$ \\ \hline
Observ. & $0.90\pm 0.05$ & $0.25\pm 0.05$ &
(4--6) & $2.2\pm 0.5$ \\ \\
\multicolumn{5}{c} {\bf $\Omega_b=0.05$} \\
  1 & 1.05 & 0.30  &  36 & 8.7\\
  2 & 1.23 & 0.34  &  53 & 11 \\
  3 & 1.23 & 0.35  &  53 & 10 \\
  4 & 0.87 & 0.19  &  19 & 1.4\\
  5 & 1.10 & 0.25  &  40 & 3.2\\
  6 & 1.23 & 0.33  &  53 & 5.1\\
  7 & 0.76 & 0.13  &  11 & 4E--02\\
  8 & 1.02 & 0.21  &  33 & 0.6\\
  9 & 1.22 & 0.36  &  52 & 2.3\\
 10 & 0.68 & 0.10  &  6.1& 4E--05\\
 11 & 0.96 & 0.18  &  28 & 2E--03\\
 12 & 1.29 & 0.53  &  60 & 0.8\\
 13 & 0.62 & 0.08  &  3.4& 1E-11\\
 14 & 0.95 & 0.21  &  27 & 1E--04\\
 15 & 1.27 & 0.59  &  58 & 0.3\\
CDM & 1.33 & 0.47  &  62 & 30 \\
CHDM& 0.86 & 0.16  &  18 & 1E--02\\ \\
\end{tabular}
\end{table}

\newpage
\thispagestyle{empty}

\begin{table}[h]
\centering
Table 2: continued \\
\tabcolsep 6pt
\begin{tabular}{ccccc}  \\ \hline \hline
Model & ${\bf \sigma_8}$ & ${\bf \Gamma}$ &
$N(>M)$ & ${\bf \Omega_g}$ \\ \hline
Observ. & $0.90\pm 0.05$ & $0.25\pm 0.05$ &
(4--6) & $2.2\pm 0.5$ \\ \\
\multicolumn{5}{c} {\bf $\Omega_b=0.08$} \\
 16 & 1.03 & 0.28  &  34 & 6.0 \\
 17 & 1.20 & 0.31  &  50 & 8.8 \\
 18 & 1.26 & 0.33  &  55 & 8.9 \\
 19 & 0.81 & 0.16  &  14 & 0.4 \\
 20 & 1.05 & 0.21  &  36 & 1.9 \\
 21 & 1.26 & 0.30  &  55 & 4.3 \\
 22 & 0.74 & 0.10  &  9.1& 3E--03\\
 23 & 1.01 & 0.18  &  32 & 0.2 \\
 24 & 1.21 & 0.35  &  51 & 1.8 \\
 25 & 0.64 & 0.08  &  3.9& 2E--07\\
 26 & 0.95 & 0.17  &  27 & 4E--03\\
 27 & 1.27 & 0.48  &  59 & 0.8 \\
 28 & 0.60 & 0.07  &  2.5& 2E-15\\
 29 & 0.93 & 0.14  &  24 & 2E--06 \\
 30 & 1.19 & 0.53  &  51 & 0.1 \\
CDM & 1.28 & 0.44  &  58 & 20 \\
CHDM& 0.82 & 0.15  &  15 &  6E--03 \\ \hline

\end{tabular}
\end{table}

\end{document}